\documentclass{aa}
\usepackage{txfonts}
\usepackage{natbib}
\usepackage{graphicx}
\usepackage{color}
\newcommand{\ft}[1]{{$^\mathrm{#1}$}}

\begin{document}

\title{Binarity of Transit Host Stars
  \thanks{Based on observations collected at the Centro Astron{\'o}mico
    Hispano Alem{\'a}n (CAHA) at Calar Alto, operated jointly by the
    Max-Planck-Institut f{\"u}r Astronomie and the Instituto de
    Astrof{\'i}sica de Andaluc{\'i}a (CSIC).
  }
}
\subtitle{Implications on Planetary Parameters}

\author{Sebastian Daemgen\inst{1}
  \and Felix Hormuth\inst{1}
  \and Wolfgang Brandner\inst{1}
  \and Carolina Bergfors\inst{1}
  \and Markus Janson\inst{1,2}
  \and Stefan Hippler\inst{1}
  \and Thomas Henning\inst{1}
}
\institute{Max-Planck-Institut f{\"u}r Astronomie, K{\"o}nigstuhl 17, 69117
  Heidelberg, Germany \and
  University of Toronto, Department of Astronomy, St. George Street 50,
  M5S 3H4 Toronto, ON, Canada\\
  \email{daemgen@mpia-hd.mpg.de}
}

\abstract
{Straight-forward derivation of planetary parameters can only be achieved in transiting planetary systems. However, planetary attributes such as radius and mass strongly depend on stellar host parameters. Discovering a transit host star to be multiple leads to a necessary revision of the derived stellar and planetary parameters.}
{Based on our observations of 14 transiting exoplanet hosts, we derive parameters of the individual components of three transit host stars (\object{WASP-2}, \object{TrES-2}, and \object{TrES-4}) which we detected to be binaries. Two of these have not been known to be multiple before. Parameters of the corresponding exoplanets are revised.}
{High-resolution ``Lucky Imaging'' with AstraLux at the 2.2\,m Calar Alto telescope provided near diffraction limited images in $i'$ and $z'$ passbands. These results have been combined with existing planetary data in order to recalibrate planetary attributes.}
{Despite the faintness ($\Delta\mathrm{mag}\!\sim\!4$) of the discovered stellar companions to TrES-2, TrES-4, and WASP-2, light-curve deduced parameters change by up to more than 1$\sigma$. We discuss a possible relation between binary separation and planetary properties, which---if confirmed---could hint at the influence of binarity on the planet formation process.}
{}
  
\maketitle

\titlerunning{Newly Discovered Transit Host Binarity}
\authorrunning{S. Daemgen et al.}

\section {Introduction}
It was not long after the discovery of the first extrasolar planet around another star \citep{may95} that the first multiple system was identified among exoplanet hosts. \citet{but97} detected 55\,Cnc, $\nu$\,And, and $\tau$\,Boo to not only host a planet, but also be bound in binaries. During the following years more and more surveys started to explore exoplanet host multiplicity resulting in the discovery of more than 30 binaries and triples among the exoplanet host sample to date \citep{mug07}.

Extensive seeing-limited studies were conducted searching for multiples among exoplanet host stars. \citet{rag06} used the Sloan Digital Sky Survey (SDSS) to detect 30 exoplanet hosts with one or more stellar companions. \citet{mug04} carried out near infrared imaging with the UFTI/UKIRT and SofI/NTT instruments in order to detect wide stellar companions. These early surveys were limited to the seeing limit, confirming theorists' claims that planetary systems in wide binaries are not greatly affected in terms of frequency and parameter statistics \citep[e.g.][]{des07}.

Very recent and ongoing high-resolution surveys \citep{egg07,cha06,mug05} started to explore smaller separations. Adaptive optics were used to resolve radial velocity (RV) planets into multiple stars where mainly the brighter component had been found to host the planet. First results show that binary statistics and parameters of planet hosts are different from those of single stars implying a modified formation, migration, and/or dynamical evolution process \citep{des07}.

Due to its currently more efficient planet detection capability, most ongoing multiplicity surveys concentrate on planet hosts found by RV. However, of today's more than 300 known exoplanets over 50 were found by photometric surveys due to their transit behavior. Transits (in combination with RV measurements) are the only way to derive the complete set of planetary parameters and therefore deserve special attention. Nonetheless, one has to be careful about parameters deduced, since one cannot rule out the possibility of a fully blended close companion to the exoplanet host star. Although a bright secondary might show up in follow-up spectra as a second set of lines, a faint ($\Delta$mag = several magnitudes) long-period companion would not be seen in the spectra as it would be buried in the primary's noise. These companions do not affect spectral parameters such as the RV semi amplitude $K_*$ and mass estimates from spectral features, but they demand another evaluation of the transit light curve resulting in a change of planet properties.

In the following we discuss the impact of binarity on transit light curve evaluation and apply this to our findings of three companions to transit hosts, namely WASP-2, TrES-2, and \mbox{TrES-4}.

\section{Sample Selection}\label{sampsel}
Our observations of 14 exoplanet hosts belong to a survey of high-resolution imaging of ultimately all transit exoplanet hosts discovered to date. The survey started in May 2007 observing the first of the $\sim\!30$ northern transit planet hosts. Due to its limiting magnitude of $i'\!\approx\!15$, all transit hosts in the northern hemisphere are accessible to our observations. The 14 exoplanet hosts that have already been observed are listed in Tables~\ref{tab1} and \ref{tab2} along with their coordinates and magnitudes.
\begin{table}
  \caption{Transiting Exoplanet Hosts Observed with AstraLux with No Stellar Companions Found\ft{\dag}}  \label{tab1}
  \begin{center}
  \begin{tabular}{lcr@{.}l}
    \hline\hline\\[-2ex]
    Name &
    2MASS Name &
    \multicolumn{2}{c}{Vmag\ft{a}} \\[0.5ex]
    \hline\\[-2ex]
    \object{HD\,209458} & J22031077+1853036 &  7&65  \\
    \object{HD\,189733} & J20004370+2242391 &  7&68  \\
    \object{HD\,17156}  & J02494447+7145115 &  8&17  \\
    \object{GJ\,436}    & J11421096+2642251 & 10&68  \\
    \object{WASP-1}     & J00204007+3159239 & 11&8   \\
    \object{TrES-1}     & J19040985+3637574 & 11&4   \\
    \object{TrES-3}     & J17520702+3732461 & 12&40  \\
    \object{HAT-P-1}    & J22574684+3840302 & 10&4   \\
    \object{HAT-P-5}    & J18173731+3637170 & 11&95  \\
    \object{HAT-P-6}    & J23390581+4227575 & 10&54  \\
    \object{XO-2}       & J07480647+5013328 & 11&25  \\[0.5ex]
  \hline
  \end{tabular}
  \end{center}
  \begin{list}{}{}
  \item[\ft{\dag}]According to our sensitivity limits (see \S\ref{sens}).
  \item[\ft{a}]V band magnitudes from Simbad  
  \end{list}
\end{table}

\begin{table*}
  \caption{Integrated Properties of the Previously Unresolved Targets}
  \label{tab2}
  \begin{center}
  \begin{tabular}{lccr@{\,$\pm$\,}lr@{\,$\pm$\,}lr@{\,$\pm$\,}lcr@{\,$\pm$\,}lc}
    \hline\hline\\[-2ex]
    &
    &
    &
    \multicolumn{2}{c}{$J$\ft{a}} &
    \multicolumn{2}{c}{$H$\ft{a}} &
    \multicolumn{2}{c}{$K$\ft{a}} &
    &
    \multicolumn{2}{c}{d} &
    \\
    Name &
    RA &
    DEC &
    \multicolumn{2}{c}{[mag]} &
    \multicolumn{2}{c}{[mag]} &
    \multicolumn{2}{c}{[mag]} &
    SpT &
    \multicolumn{2}{c}{[pc]} &
    Ref.     \\[0.5ex]
    \hline\\[-2ex]
  \object{WASP-2} & 20\,30\,54.13 & +06\,25\,46.37 & 10.17&0.03 &  9.75&0.03 &  9.63&0.02 & K1V  & 147 & 17\ft{b} & 1,2\\
  \object{TrES-2} & 19\,07\,14.04 & +49\,18\,59.07 & 10.23&0.02 &  9.92&0.03 &  9.85&0.02 & G0V & 213&11 & 3,4\\
  \object{TrES-4} & 17\,53\,13.06 & +37\,12\,42.36 & 10.58&0.02 & 10.35&0.02 & 10.33&0.02 & F8\ft{c} & 485&31 & 4\\[0.5ex]
  \hline
  \end{tabular}
  \end{center}
  \begin{list}{}{}
  \item[\ft{a}]Magnitudes are from the 2 Micron All Sky Survey (2MASS) catalogue \citep{skr06}.
  \item[\ft{b}]{The distance of WASP-2 has been derived indirectly from V band magnitudes (see \S\ref{phot}).}
  \item[\ft{c}]Since there are no good values published for TrES-4, we estimate the SpT from the $J-H$ color using templates by \citet{all99}.
  \item[References.] (1) \citealt{cam07}; (2) \citealt{all99}; (3) \citealt{soz07}; (4) \citealt{tor08}.
  \end{list}
\end{table*}

\section{Observations and Data Reduction}\label{obs}
\subsection{Direct Imaging with AstraLux at Calar Alto}\label{alux}
Observations were obtained with the {\em AstraLux} instrument attached to the Cassegrain focus of the 2.2\,m telescope at Calar Alto observatory in the time from May 2007 to November 2007. Reobservations of TrES-4 were conducted in June 2008, since first epoch observations have been incomplete. The instrument employs the ``Lucky Imaging'' technique \citep{law06} which uses several thousand short ($\sim\! 10\,$ms) exposures in order to minimize the effect of atmospheric seeing, coadding only the least distorted images for further reduction. Owing to Lucky Imaging, diffraction limited images can be obtained without the need for adaptive optics \citep[and references therein]{hor08}.

The {\em AstraLux} instrument consists of an electron multiplying CCD which produces images with a pixel scale of 46.6\,mas/pixel which is resampled to 23.3\,mas/pixel through drizzle combination of the best 2.5, 5, and 10\,\% of the raw images respectively. The full width at half maximum (FWHM) of the final Point Spread Functions (PSF) are on average 0\farcs1, drawing level with adaptic optics observations at similar sized telescopes. 

A more extensive introduction to {\em AstraLux} can be found in \citet{hor08}.

\subsection{Photometry \& Relative Astrometry}\label{phot}
\begin{figure*}
  \centering
  \includegraphics[angle=0,width=17cm]{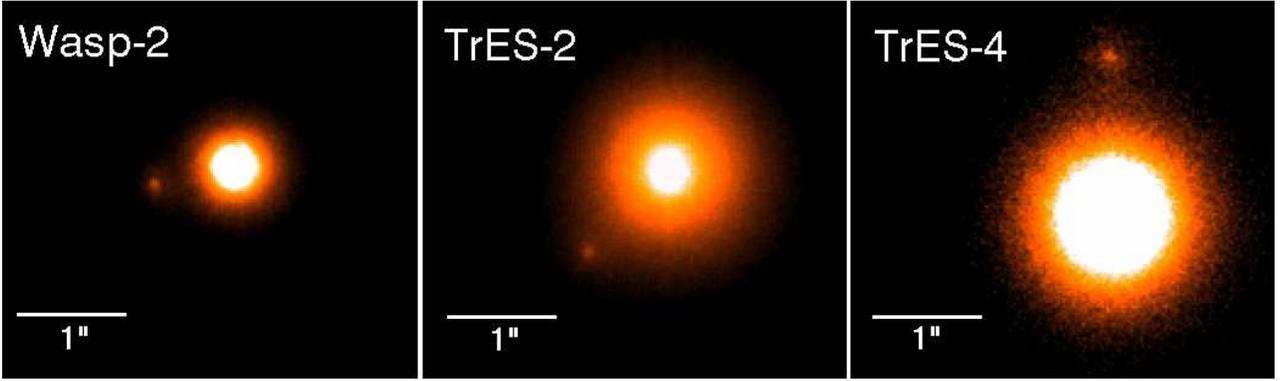}
  \caption{$z'$ filter images of the three exoplanet hosts with {\em AstraLux} at the Calar Alto observatory 2.2\,m telescope. Total integration times are 7.68\,sec for WASP-2, 15.36\,sec for TrES-2, and 11.52\,sec for TrES-4. The image scale is linear and cuts are adapted to best depicture the secondary. North is up and East is to the left.}
    \label{fig1}
\end{figure*}

Final $i'$ and $z'$ images were compiled by use of a pipeline which selects the individual images with the highest Strehl ratios and runs the shift-and-add drizzle combination. The final images clearly show a companion to three of our targets (WASP-2, TrES-2, and TrES-4) in separations that place them within the primary's PSF wings. Images of the binaries taken in $z'$-band are shown in Fig.~\ref{fig1}, coordinates and magnitudes as well as integrated spectral types (SpT) and distances are listed in Table~\ref{tab2}.
{Since no uncertainty estimates are published for the distance of WASP-2, we derive its distance from apparent \citep{str07} and absolute $V$ magnitudes from the SpT--$M_V$ relation in \citet{all99}. Uncertainty is estimated from derivation of distance values from upper and lower SpT assuming an uncertainty of one sub class. The value in Table~\ref{tab2} of $147$\,pc matches earlier determinations and the uncertainty seems to be in a reasonable range.}

Separations and position angles of those binaries were determined using the ALLSTAR PSF-fitting task in IRAF. For each target an individual PSF was built the following way: We determine the size of a rectangle sufficiently sized to contain almost all of the faint companion's flux ($\sim$\,50\,$\times$\,50\,px $\approx\!1\farcs2\times1\farcs2$). The content of a box of the same size was then copied from the opposite side of the primary's flux maximum. Rotating this box by 180$^\circ$ and shifting it to the secondary's position, the flux distribution within the box nicely fits into the surrounding flux distribution replacing the secondary's signature. The resulting PSF consists of the primary's PSF without light from the secondary. In some images---where the PSF decisively deviates from a point symmetric shape---residual ``steps'' at the edges of the inserted box are visible. {However, step sizes never exceed 20\% of the surrounding total flux count and all discontinuities are well separated from the PSF's core. The central region of the PSF, which dominates the photometry results, remains unchanged.} The accuracy of this method was tested using PSF stars without contamination by a close companion. Comparing photometry on several objects with and without copying a region according to the above scheme, we did not measure any differences in magnitude and location (nominal display precision $\delta\mathrm{mag_{min}}\!=\!0.001$). Hence, the accuracy of deduced results is only negligibly influenced by this method.

This rather unconventional approach to finding a suitable PSF is preferable over PSF stacking from other observations, as the {\em AstraLux} PSF is subject to variations and additional bright sources suitable as reference PSF's have been found in only very few of the science acquisitions. Those have not been used for photometry in order to consistently evaluate the photometric data.

PSF photometry yields stellar locations with subpixel accuracy. From this, separation and position angle were determined. Astrometric and photometric data are summarized in Table~\ref{tab3}.

\begin{table*}
  \caption{Observed Binary Properties.}
  \label{tab3}
  \centering
  \begin{tabular}{lr@{\,$\pm$\,}lr@{\,$\pm$\,}lr@{\,$\pm$\,}lr@{\,$\pm$\,}lc}
    \hline\hline\\[-2ex]
    &
    \multicolumn{2}{c}{$\Delta i'$} &
    \multicolumn{2}{c}{$\Delta z'$} &
    \multicolumn{2}{c}{sep} &
    \multicolumn{2}{c}{PA} &
    Date \\
    Name &
    \multicolumn{2}{c}{[mag]} &
    \multicolumn{2}{c}{[mag]} &
    \multicolumn{2}{c}{[\arcsec]} &
    \multicolumn{2}{c}{[$^\circ$]} &
    (UT) \\[0.5ex]
    \hline\\[-2ex]
    WASP-2 & 4.095&0.025 & 3.626&0.022 & 0.757&0.001 & 104.7&0.3 & Nov 2007\\
    TrES-2 & 3.661&0.016 & 3.429&0.010 & 1.089&0.008 & 135.5&0.1 & May 2007\\
    TrES-4 & 4.560&0.017 & 4.232&0.025 & 1.555&0.005 & 359.8&0.1 & Jun 2008\\[0.5ex]
  \hline
  \end{tabular}
\end{table*}

\section{Stellar Properties}
\subsection{Spectral Types, Magnitudes, \& Distances}\label{spt}
{For WASP-2 and TrES-2 published spectral types of the unresolved binaries are available. However, published spectral types for TrES-4 seem to be inconsistent. We estimate the SpT of TrES-4 from its $J-H$ color using templates by \citet{all99}. The resulting SpT of F8 will be used in this paper.} In order to determine separate SpTs of the A and B component respectively, integrated SpTs as well as integrated and separate colors were used.

The SpT of the primary component is assumed to be identical to the integrated SpT that was observed without the knowledge about the binarity of the target (see Table~\ref{tab2}). This is a reasonable assumption due to the faintness of the companion. As SpTs of all three objects were determined by the means of spectroscopy, a $\Delta\mathrm{mag}\!\approx\!4$ secondary does not influence the SpT determination since it is not visible in the spectra. Secondary component SpTs, however, need to be estimated from their color.

\begin{figure}
  \centering
  \includegraphics[angle=0,width=\columnwidth]{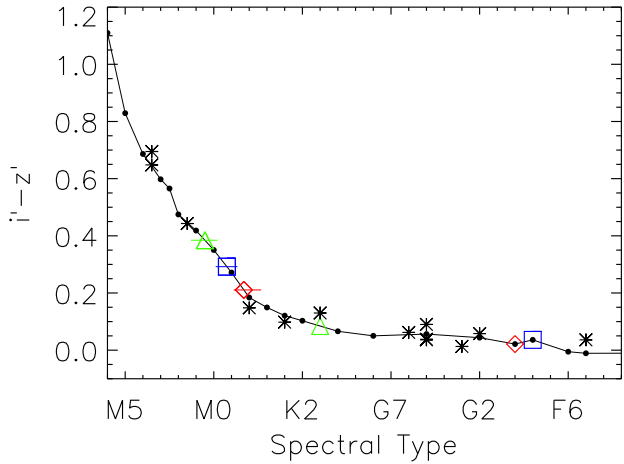} 
  \caption{$i'-z'$ color vs. spectral type derived from template spectra from \citet{pic98} convolved with the sdss $i'$ and $z'$ filter curves. Overplotted are the best values for A and B component of three objects: \textcolor{green}{$\triangle$} (WASP-2), \textcolor{red}{$\diamond$} (TrES-2), and \textcolor{blue}{$\square$} (TrES-4). The bars on the three leftmost items denote the spectral uncertainty of the B component of each of the three binaries, respectively. Since we assume a SpT from the literature for the A component no error bar is given. Asterisks ($\times$\hspace{-1.42ex}$+$) denote the positions of SDSS standard main sequence stars from \citet{smi02}. These are included to verify the accuracy of the found relation as well as provide an uncertainty estimate of colors derived.}
    \label{fig2}
\end{figure}

A color--SpT diagram ranging from F5 to M5 was composed from template spectra \citep{pic98} which were convolved with filter curves of the SDSS $z$ and $i$ filters at zero airmass\footnote{http://www.sdss.org/dr3/instruments/imager/index.html} (Fig.~\ref{fig2}).
Zeropoints were determined according to the definition of AB magnitudes \citep{oke74}. The small correction to convert from SDSS $i$, $z$ to $i'$, $z'$ magnitudes was applied according to the formula given on the SDSS website. Comparing our synthetic $i'-z'$ values to those of SDSS main sequence standard stars \citep[from][see Fig.~\ref{fig2}]{smi02}, we find that the relation very well fits observed data. We use the $\delta (i'-z')=0.033$ scatter of standard colors about our color-SpT curve as a measure for the $i'-z'$ uncertainties from our relation. This uncertainty applies to all primary component $(i'-z')_A$ colors which are derived from SpT and our color-SpT relation. {The color of the secondary component is calculated through $(i'-z')_{/C} = \Delta i'-\Delta z'-(i-z)_A$ with $\Delta i'=i'_{/C}-i'_A$ and $\Delta z'= z'_{/C}-z'_A$ from Table~\ref{tab3}, and $(i'-z')_A$ is taken from the A components' values in Table~\ref{tab4}.}

Reading off spectral types from the $(i'- z')$--SpT relation using our separate $i'- z'$ colors (see Table~\ref{tab4}), we receive individual SpT values with uncertainties of about 1 spectral class for SpTs later than $\sim$\,K5.
\begin{table*}
  \caption{Individual Binary Component Parameters.}
  \label{tab4}
  \begin{center}
    \begin{tabular}{lccr@{\,$\pm$\,}lccr@{\,$\pm$\,}lr@{\,$\pm$\,}lr@{\,$\pm$\,}lc}
      \hline\hline\\[-2ex]
      &
      &
      $M_\ast$ &
      \multicolumn{2}{c}{$(i'-z')$} &
      $i'$\ft{b} &
      $z'$\ft{b} &
      \multicolumn{2}{c}{$M_{i'}$} &
      \multicolumn{2}{c}{$d_\mathrm{phot}$} &
      \multicolumn{2}{c}{sep\ft{c}} &
      $P$\ft{c,d} \\
      Name\ft{a} &
      SpT &
      [$M_\odot$] &
      \multicolumn{2}{c}{[mag]} &
      [mag]&
      [mag]&
      \multicolumn{2}{c}{[mag]} &
      \multicolumn{2}{c}{[pc]} &
      \multicolumn{2}{c}{[AU]} &
      [$10^3$ yr] \\[0.5ex]
      \hline\\[-2ex]
      WASP-2A  & K1              &0.77&  0.084&0.033 & 11.27 & 11.19 & 5.74&0.19 & 127&13 & 111    &13    & 1.4        \\
      WASP-2/C & M0\,\dots\,M1.5 &0.48&  0.385&0.047 & 15.38 & 14.81 & 7.95&0.29 & 310&40 & \multicolumn{2}{c}{}&   \\
      TrES-2A  & G0              &1.05&  0.021&0.033 & 11.07 & 11.04 & 4.45&0.11 & 211&14 & 232    &         12 & 3.9           \\
      TrES-2/C & K4.5\,\dots\,K6 &0.67&  0.211&0.038 & 14.73 & 14.47 & 6.73&0.20 & 400&40 & \multicolumn{2}{c}{}&    \\
      TrES-4A  & F8              &1.18&  0.036&0.033 &  9.33 &  9.29 & 4.27&0.18 & 103&10 & 755    &49          & 21  \\
      TrES-4/C & K5.5\,\dots\,M0 &0.59&  0.292&0.045 & 13.85 & 13.57 & 7.35&0.54 & 200&50 & \multicolumn{2}{c}{}&    \\[0.5ex]
      \hline
    \end{tabular}
  \end{center}
  \begin{list}{}{}
  \item[\ft{a}]{The stellar secondary is referred to as '/C' in order to avoid confusion with the notation of the planetary companion, which is referred to as an appendix 'b' to the host star's name in many publications.}
  \item[\ft{b}]{Uncertainties of apparent magnitudes are estimated to be $\delta\rm{mag}=0.10$ (see \S\ref{spt}).}
  \item[\ft{c}]{Separation and period estimates use distances from Table~\ref{tab2}.}
  \item[\ft{d}]{{Periods are order-of-magnitude estimations and hence do not include an uncertainty estimation.}}
  \end{list}
\end{table*}
Individual uncertainties were estimated by determining the SpT from the curve for upper and lower limit of the color measurement respectively. Spectral types earlier than K5 are less reliably determined from the diagram due to the rapidly dropping gradient of $i'-z'$ color with SpT. However, all secondary's spectral types are within or close to the steep region of SpT later than K5. Hence, SpTs are believed to be reasonably well determined by this method.

Since there are no published $i'$ and $z'$ magnitudes to be found for the unresolved exoplanet hosts we estimate separate $i'$ and $z'$ magnitudes of A and B components from photometry of reference objects. We use images of the SDSS standards SA\,112-223, Ross\,711, and BD+71\,0031 which we observed the during same night at similar airmass as WASP-2, TrES-2, and TrES-4 respectively. This enables us to establish zero points of our $i'$ and $z'$ exposures by comparing aperture photometry results from the {\it IDL} {\it ATV}\footnote{http://www.physics.uci.edu/\~{}barth/atv/} routine to SDSS standard photometry. These zeropoints are then applied to the science targets' magnitude measurements.

Apparent magnitude uncertainties are estimated to be $\delta \mathrm{mag}\!\approx\!0.10$ for the following reasons. Aperture photometry allows a nominal precision of $\approx\!0.03$\,mag which is derived from the change of measured magnitude when varying the three aperture photometry parameters (aperture radius, annulus radius, \& annulus width) within a reasonable range. Converting these numbers into zeropoints and absolute magnitudes this uncertainty piles up to about 0.07 in some cases. Furthermore, exposures are not corrected for airmass, but taken at similar airmasses. We account for that by assuming an additional uncertainty. Hence, total uncertainties of our absolute photometry are conservatively assumed to be $\delta \mathrm{mag}\!\approx\!0.10$. The results are tabulated in Table~\ref{tab4}.

Distances are estimated photometrically from apparent and absolute $M_{i'}$ magnitudes. The latter are derived from our SpT estimates and a SpT--$M_{i}$ relation by \citet{kra07} which was converted to $M_{i'}$ applying the SDSS conversions. $M_{i'}$ values were derived applying cubic spline interpolation in order to reach values that are not explicitely published in their table. Since there are no uncertainties published for the $M_{i'}$ curve, we estimate uncertainties for the A components by assuming a SpT uncertainty of one subclass. For the B component we use the spectral uncertainty in Table~\ref{tab4}. We derive and compare distances for individual components to decide whether the binaries are physical (see \S\ref{phys}).

Individual component parameters are summarized in Table~\ref{tab4}.

\subsection{Masses \& Periods}\label{mass}
None of the stellar secondaries were individually investigated by use of a spectrograph. Hence, deduction of mass from spectral features has not been possible for the separate components. Masses of the individual binary components are estimated according to the following, assuming physical attachment of the binary components. A discussion of physical binarity can be found in \S\ref{phys}.

Since the secondary is not observable in the acquired spectra, we assume spectral measurements to correspond to the primary's properties. Hence, characteristics like e.g. mass derived from stellar models \citep[see][for a consistent discussion of all three objects]{tor08} need to be attributed to the primary.

Our calculations rely on numbers by \citet{tor08} who collected host mass and radius through comparison of 
\begin{equation}\label{eq1}
  \left(\frac{a}{R_*}\right)_{Y^2} = \left(\frac{G}{4\pi^2}\right)^{1/3}\frac{P^{2/3}}{R_*}\left(M_*+M_\mathrm{p}\right)^{1/3}
\end{equation}
with input from Yonsei-Yale (Y$^2$) Models \citep{yi01} and 
\begin{equation}
   \left(\frac{a}{R_*}\right)_\mathrm{transit} = \left\{\frac{(1+R_\mathrm{p}/R_*)^2-b^2[1-\sin^2(t_T\pi/P)]}{\sin^2(t_T\pi/P)}\right\}^{3/2}
\end{equation}
entirely fed with light curve parameters. In this paper we derive updated values of $(a/R_*)_\mathrm{transit}$ taking the binarity of the transit host star into account (see \S\ref{new}). However, we expect negligible impact on the deduced primary mass when adapting the models in order to match the new $a/R_*$ value: newly derived $a/R_*$ are 0.9\%, 1.0\%, and 0.5\% bigger than the previously reported values for WASP-2, TrES-2, and TrES-4 respectively. {This translates to a somewhat larger change in mass driven by the $1/3$ exponent of the masses in eq.~(\ref{eq1})}: a 1\% increase of $a/R$ results in a 3\% increase of stellar mass. This is clearly within the uncertainty of the mass estimations of any of the three targets which are on the order of 5--12\%. We therefore adopt the mass that was previously attributed to the formerly supposed single star as the primary mass of the newly discovered binary.

In order to estimate a binary period by the use of the 3rd Kepler law \citep{kep1619}, secondary masses are estimated from our SpT according to \citet{all99} (see Table~\ref{tab4}). However, the numbers used for estimation of periods in Table~\ref{tab4} are rough estimates only since masses are calculated for zero age main sequence and do not include the effect of metallicity. Hence, if better values were known, those would have been used for calculations, as is the case for all primary components. Their spectroscopically determined masses are listed in Table~\ref{tab5}.
\begin{table*}
  \caption{{Literature Values used to Calculate the New Parameters}.}
  \label{tab5}
  \begin{center}
  \begin{tabular}{c@{\dotfill}cr@{\,$\pm$\,}lr@{\,$\pm$\,}lr@{\,$\pm$\,}l}
  \hline\hline
    \hspace{1cm}Parameter\hspace{1cm} &
    &
    \multicolumn{2}{c}{WASP-2}&
    \multicolumn{2}{c}{TrES-2}&
    \multicolumn{2}{c}{TrES-4} \\[0.5ex]
  \hline
    $\Delta F$\ft{a}    &&0.01713 &0.00020 &0.01570&0.00013&9.806941&0.000087 \\
    $P$ (days)          &&2.152226&0.000004&2.47063&0.00001&3.553945&0.000075 \\
    $t_{IV}-t_{I}$ (hr) &&   1.799&   0.035&  1.840&  0.020&   3.638&  0.0824\ft{b}\\
    $t_{II}-t_{I}$ (hr) &&    0.41&    0.04&  0.683&  0.045&   0.701&   0.026\ft{b}\\
    $K$ (m\,s$^-1$)     &&     155&       7&  181.3&    2.6&    97.4&     7.2 \\
    $T_\mathrm{eff}$ (K)&&    5200&     200&   5850&     50&    6200&      75 \\
    $M_\mathrm{*}$ (M$_\odot$)&&0.89&0.12&0.983&0.061&   1.394&   0.058 \\
    Ref.&&\multicolumn{2}{c}{1,2}&\multicolumn{2}{c}{3,4}&\multicolumn{2}{c}{5,6}\\[0.5ex]
  \hline
  \end{tabular}
  \end{center}
  \begin{list}{}{}
\item[\ft{a}]{$\Delta F=(F_\mathrm{no\,transit}-F_\mathrm{transit})/F_\mathrm{no\,transit}=(R_\mathrm{p}/R_*)^2$}
\item[\ft{b}]{Transit duration numbers for TrES-4 are inferred from values published by \citet{man07} using inverted formulas from \citet{sea03}.}
\item[References.]{(1) \citealt{cam07}; (2) \citealt{cha07}; (3) \citealt{odo06}; (4) \citealt{hol07}; (5) \citealt{man07}; (6) \citealt{tor08}.}
\end{list}
\end{table*}
To account for random inclinations and eccentricities, projected separations have been multiplied by a factor of 1.26 according to simulations by \citet{fis92}. Period estimates and separations are listed in Table~\ref{tab4}.

\section{New System Parameters of Planet and Host}\label{new}
We derive a number of parameters of the planets as well as their host stars. Since earlier calculations did not take the blending with the additional sources into account, system parameters will change in respect to previously published numbers.

Transit light curves offer the great opportunity of making fundamental planetary parameters directly accessible. Mass, radius, and other planet properties can be calculated from the four light curve parameters and one additional stellar parameter. The light curve parameters are: Period $P$, transit duration $t_{IV}-t_I$, ingress/egress duration $t_{II}-t_I$, and dip depth $\Delta F=(F_\mathrm{no\,transit}-F_\mathrm{transit})/F_\mathrm{no\,transit}$. From these values a number of planetary and stellar parameters can be derived, these are: $R_\mathrm{p}/R_\ast$, $a/R_\ast$, impact parameter $b$, inclination $i$, and stellar density $\rho_\ast$ \citep[cf.][]{sea03}, where $R_\mathrm{p}$ and $R_\ast$ are the radii of planet and star respectively and $a$ is the semimajor axis of the planet orbit. To resolve above ratios of system parameters, one additional piece of information is needed. This could be either stellar mass or stellar radius. We choose to use $M_\ast$ since stellar radius is typically less well constrained by stellar models. If we assume stellar mass from the literature, stellar radius can be inferred from $\rho_\ast$ which is known from the light curve parameters.

Additional parameters can be derived with knowledge of the radial velocity amplitude $K_\ast$ and the effective temperature $T_\mathrm{eff}$ of the host star. Both numbers can be obtained from spectral measurements which have been acquired for all confirmed transiting planet hosts to date.

\subsection{Impact of Host Binarity on Stellar \& Planetary Parameters}\label{imp}
The evaluation of the transit light curve generally assumes host singularity. Subtracting the constant flux offset of a blended binary component, $\Delta F$ will assume a greater value than $\Delta F_\mathrm{old}$ which has been calculated under disregard of an existing blend. The new $\Delta F$ is derived through
\begin{equation}
   \Delta F=(1+10^{-\frac{\Delta z^\prime}{2.5}})\cdot \Delta F_\mathrm{old}
\end{equation}
where $\Delta z^\prime$ is the magnitude difference between the primary and the secondary stellar component of the exoplanet host in SDSS $z'$ Filter (see Table~\ref{tab3}). 

The formula assumes the brighter component to be the variable component. Provided that the fainter companion would be the variable part, an eclipsing binary would be---at first sight---a valid solution. This, however, has been ruled out by line-bisector analysis for all three objects \citep{cam07,odo06,man07}. Additionally, in the case of WASP-2 the necessary $\sim$1.5\,mag eclipse of the secondary in $H$ band in order to explain the light curve could not be observed \citep{cam07}.

For further evaluation, we use formulae (6) through (9) in \citet{sea03} which analytically connect $R_\mathrm{p}/R_*$, $b$, $a/R_*$, and $\rho_*$ to light curve parameters. Further system parameters (planetary surface gravity $\log g_\mathrm{p}$, Safronov number $\Theta$, and equilibrium temperature $T_\mathrm{eq}$) are derived according to the formulae given in \citet{tor08}.

{Reported uncertainties are copied from published values in the observers' papers. Since the updated numbers in this paper differ by less than 2\% from the old ones, the uncertainties are expected to change in a similar manner on a scale of not more than a few percent. We therefore keep the previous uncertainties since our computational approach cannot reproduce uncertainties with all observational constraints that have been included in the original papers.}

Input values used are assembled in Table~\ref{tab5}. The resulting set of new parameters is listed in Table~\ref{tab6}.

\begin{table*}
  \caption{The New System Parameters}
  \label{tab6}
  \begin{center}
  \begin{tabular}{c@{\dotfill}cr@{\,$\pm$\,}lr@{\,$\pm$\,}lr@{\,$\pm$\,}l}
  \hline\hline
    \hspace{1cm}Parameter\hspace{1cm} &
    &
    \multicolumn{2}{c}{WASP-2}&
    \multicolumn{2}{c}{TrES-2}&
    \multicolumn{2}{c}{TrES-4}\\[0.5ex]
  \hline
    $R_\mathrm{p}/R_*$           & & 0.1332&0.0015 & 0.1279&0.0010 & 0.1000&0.0009\\
    $a/R_*$                      & &   8.01&$^{0.32}_{0.2}$&   7.65&0.12   &   6.05&0.13  \\
    $b$                          & &  0.725&0.026  &  0.851&0.006  &  0.752&0.015 \\
    $i$ $(\mathrm{deg})$         & &  84.80&0.39   &  83.62&0.14   &  82.86&0.33  \\
    $\rho_*$ $(\mathrm{g/cm^3})$ & &  2.095&0.205  &  1.390&0.060  &  0.333&0.022 \\
    $\log g_\mathrm{p}$          & &  3.279&0.036  &  3.284&0.016  &  2.867&0.038 \\
    $R_*/R_\odot$                & &  0.843&0.063  &  0.999&0.033  &  1.809&0.064 \\
    $R_\mathrm{p}/R_\mathrm{Jup}$& &  1.117&0.082  &  1.272&0.041  &  1.799&0.063 \\
    $M_\mathrm{p}/M_\mathrm{Jup}$& &  0.914&0.092  &  1.199&0.052  &  0.919&0.073 \\
    $\rho_\mathrm{p}$ $(\mathrm{g/cm^3})$& &   0.87&0.24   &   0.77&0.09   &  0.209&0.029 \\
    $\Theta$                     & & 0.0590&0.0044 & 0.0697&0.0022 & 0.0381&0.0030\\
    $T_\mathrm{eq}$ $(\mathrm{K})$& &   1300&54     &   1495&17     &   1782&29    \\
    $a$ $(\mathrm{AU})$          & &0.03138&0.00142&0.03556&0.00075&0.05091&0.00071\\[0.5ex]
  \hline
  \end{tabular}
  \end{center}
\end{table*}

\subsection{Notes on the individual Objects}
\subsubsection{WASP-2}
WASP-2b is the second bona fide transiting planet detected by the SuperWASP survey \citep{pol06} and followed up with the SOPHIE spectrograph \citep{bou06} at the Observatoire de Haute Provence. The discovery paper \citep{cam07} presents a $R_\mathrm{p}=(0.65-1.26)$\,$R_\mathrm{Jup}$, $M_\mathrm{p}=(0.81-0.95)$\,$M_\mathrm{Jup}$ planet on a 2.15 day orbit around \mbox{WASP-2}.

\citet{cam07} report a stellar companion to WASP-2 $0\farcs7$ to the East which we confirm with our finding of a $\Delta i'=4.095\pm0.025$, $\Delta z'=3.626\pm0.022$ companion at a consistent position. However, removing the secondary light from the light curve evaluation {requires us} to augment the former value of $R_\mathrm{p}/R_*$ by 1.76\%, which is greater than the absolute uncertainty of 0.0015\,mag. This results in a 0.9\% increase of absolute planet radius $R_\mathrm{p}$.

\subsubsection{TrES-2}
This transit host with the alternative identifier GSC\,03549-02811 has been discovered in 2006 by the Trans-Atlantic Exoplanet Survey \citep[TrES;][]{odo06}. Due to the smallness of reported uncertainties in the discovery paper and a later refinement of parameters by \citet{hol07}, the relative change of parameters in terms of significance is the biggest for the \mbox{TrES-2} system. We calculate a 2.1\% increase of $R_\mathrm{p}/R_*$ which is more than twice the uncertainty of its absolute value. Other parameters change up to 0.7$\sigma$ with respect to previously derived values.

\subsubsection{TrES-4}
The transiting planet TrES-4b is worth mentioning for several reasons. Having been discovered by \citet{man07} to orbit the late F star GSC\,02620-00648 it was classified to be the least dense of all transiting planets known at the time of writing, due to its large radius of $R_\mathrm{p}=1.80\pm0.06 R_\mathrm{Jup}$ (this paper). Similarly, a low value for the parent star's density is calculated. Furthermore, TrES-4 is the only object in our 3-object sample that belongs to the ``Class II'' objects of low Safronov numbers $\Theta$ at high equilibrium temperatures $T_\mathrm{eq}$ \citep{han07}. Implications to draw from $\Theta$ are discussed in \S\ref{saf}.

\section{Discussion}\label{dis}

\subsection{Sensitivity Limit to the Detection of Stellar Companions}\label{sens}
Table~\ref{tab1} lists all observed transiting planet hosts observed in our 2007 run that do not show signs of stellar companions. We can exclude companions according to our sensitivity limits discussed below.

The dim end of detectable point-like sources is determined by the resolution and background noise of our {\em AstraLux} observations. With an average FWHM of the PSF of $\sim\!0\farcs1$, the detection of sources several arcseconds away from a bright source is limited by the detection limit of $\Delta i'\!\lesssim\!8$ of our {\em AstraLux} observations.

In proximity (comparable to the FWHM of $\sim\!0\farcs1$) to a star, however, the detection limit is dominated by the photon noise of the primary source. Hence, the minimum detectable magnitude is the greater the tighter the binary is. Since deriving binary statistics is not the primary goal of this paper, we do not develop a thorough sensitivity examination for very tight binaries and discontinue further evaluation here.

We would like to note that close-in bright companions would have shown up in previously observed spectra. From this constraint a limiting magnitude could be formulated. However, this is dependent on the spectrograph used and hence cannot be discussed in detail here. For binaries with separations greater or comparable to the FWHM, this limit is less strong than our detection limit which pushes the mass down to or even below the deuterium burning limit. The non-detection of binaries from spectroscopy, however, excludes binaries with projected separations tighter than possible to resolve with imaging techniques.

For reasons of a limited field of view of $\sim\!12\arcsec\!\times\!12\arcsec$ and in order to be complementary to previous observations we chose to restrict binary detections to separations $\lesssim\!2\arcsec$. Wide binaries separated by several arcseconds would have been resolved even with lower than near diffraction-limited resolution; seeing-limited observations have been done for all confirmed transiting exoplanet hosts. In the course of those observations it has been found that two objects, HD\,189733 and HAT-P-1, have wide companions \citep{bak06,bak07}. Since separations of both objects are more than 10$\arcsec$ they could not have been detected with our instrumental setup nor would they have passed our separation selection criterion. Nevertheless we include those two findings into our analysis in \S\ref{saf}, since they helpfully enlarge the sample of transit host binaries in order to draw more reliable statistical conclusions.

\subsection{Are the Companions Physically Bound?}\label{phys}
\citet{kra07} present absolute magnitudes in the SDSS $ugriz$ filter system allowing to derive {$M_{u'}M_{g'}M_{r'}M_{i'}M_{z'}$} absolute magnitudes (by applying the conversion formulae) and hence distances for each of the individual target components. Individual distances are used to test whether the components reside in the same distance to Earth and are therefore physically related. Results (Table~\ref{tab4}) seem to show a discrepancy between each component pair, for example \mbox{WASP-2A} is calculated a distance of $127\!\pm\!13$\,pc whereas WASP-2/C is suggested to reside in $310\!\pm\!40$\,pc distance. However, distance uncertainties are probably underestimated due to missing absolute magnitude uncertainties from \citet{kra07} who derive their $M_{i'}$ values from a multitude of color-SpT and color-color relations.

Since all three objects show a secondary distance about two times the primary distance, systematic errors might play a role. For example, the absolute photometry assumes the objects to reside on the main sequence. TrES-4, with the biggest discrepancy between $V$ band photometric distance (Table~\ref{tab2}) and distance derived from $i'$ magnitudes (Table~\ref{tab4}), was suggested by \citet{man07} to have slightly evolved from the main sequence towards the subgiant branch, which would explain the high $i'$ magnitude leading to the small photometric distance. Hence, although photometric distances suggest non-related binary components, we cannot finally decide physical binarity.

{A statistical approach can be achieved by estimating the density of background sources and deriving the expected number of targets accompanied by a background giant within $2\arcsec$. Since the sky coverage of the SDSS catalog is not sufficient to cover all targets in the sample, we used the 2MASS catalog in order to find all sources within $30\arcmin$ of each target. The density of giants $\rho(m_K)$ detectable with AstraLux can be estimated with the following cuts. Colors of $J-K\ge0.5$ are selected to include mostly giants while the limiting magnitude of AstraLux translates to roughly $m_K\approx14$. Applying these cuts we find the probability to detect a background source brighter than $m_K$ within a separation of $\Theta$ according to \citep[see][]{bra00}
\begin{equation}\label{eq4}
  P(\Theta,m_K) = 1-e^{-\pi \rho (m_K) \Theta^2}\quad.
\end{equation}
The average probability to find one or more giants close to a target is $P=0.16$\% according to eq.~\ref{eq4}. Hence, the expected number of sources with non-related background companions is $E=0.022$, which strongly suggests that the three companion sources in this sample are of physical nature.}

\citet{cam07} report a stellar companion to \mbox{WASP-2} that is likely to be identical to our observed companion. Their high-resolution observations in September 2006 indicate a $\Delta H=2.7$ companion located $0\farcs7$ to the East which agrees with our observed PA of $(104.7\pm 0.3)^\circ$ and separation of $0\farcs757\pm 0\farcs001$ in November 2007. WASP-2, with its proper motion of 53.2\,mas/yr \citep{zac04} and a time baseline of $\sim\!1$\,yr, is expected to show a significant change in separation if the components were unrelated. However, since there are no uncertainties published we cannot definitively state that WASP-2 and its faint companion share proper motion.

Despite above considerations, a reliable confirmation or rejection of physical binarity can only be achieved by second epoch data with a sufficiently long time baseline. 

We would like to point out that physical binarity is irrelevant for derivations in \S\ref{new}, \S\ref{sens}, and \S\ref{par} which discuss the impact of blending on planetary and stellar parameters. Calculations depend on the contamination by sources in small angular separation to the target, regardless of their actual distance.

\subsection{The New Parameters}\label{par}
In the following we will refer to values that were derived without the knowledge about the host star's binarity as ``old'', whereas our newly derived values from which the secondary contribution has been removed are referred to as ``new''.

The change in $\Delta F$ propagates through all numbers derived from the light curve and causes changes of up to more than 3\%. Quantifying the change in terms of significance, we find deviations of the new values compared to the old ones of up to more than 2$\sigma$ for the value of $R_\mathrm{p}/R_*$. Other parameters (listed in Table~\ref{tab6}) needed to be revalued by up to 1$\sigma$.

{The non-random nature of this effect makes deviances significant although they are 2$\sigma$ and less. For example, in the presence of a companion, $R_\mathrm{p}/R_*$ can only be greater than assumed before. The size of augmentation is determined by the brightness of the companion. A brighter companion causes a larger change of the parameter. The new parameters hence imply a shift of the mean value.}

{The relative uncertainties of the TrES-2 parameters are smaller than for the other two targets. This makes its characteristics subject to the most significant revisions.}

\subsection{Planets in the Binary Environment}\label{saf}
In an attempt to identify correlations between various planetary parameters and host star properties, \citet{han07} also compute the equilibrium temperature and Safronov number, which is defined as $\Theta = (a/R_\mathrm{p})(M_\mathrm{p}/M_*)$, {for a sample of 20 transiting exoplanets. They identify two classes of exoplanets based on their Safronov numbers. Class I planet have $\Theta = 0.07 \pm 0.01$ and Class II planets $\Theta = 0.04 \pm 0.01$. This distinction into two classes of exoplanets is confirmed by \citet{tor08a}, who increased the sample to 23 sources, and also took stellar evolution into account when computing the physical properties.}
\citet{han07}  speculate that the two classes might be the outcome of different migration mechanisms. Another possibility discussed is that a decrease in the average density due to mass loss by selective evaporation of of Helium could explain the presence of Class II planets with the abnomally large radii.

\begin{figure*}
  \centering
  \includegraphics[angle=0,width=\columnwidth]{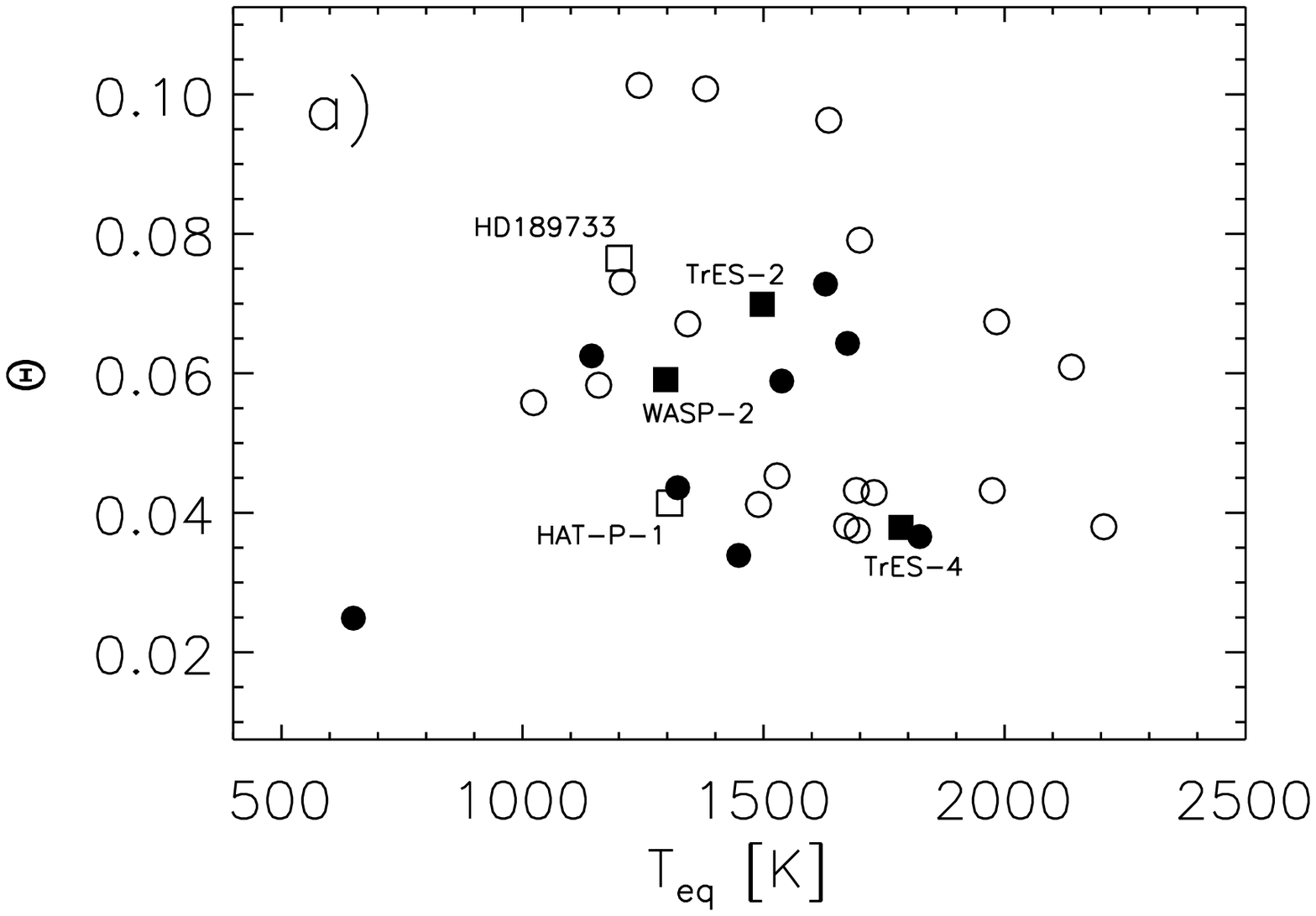}
  \includegraphics[angle=0,width=\columnwidth]{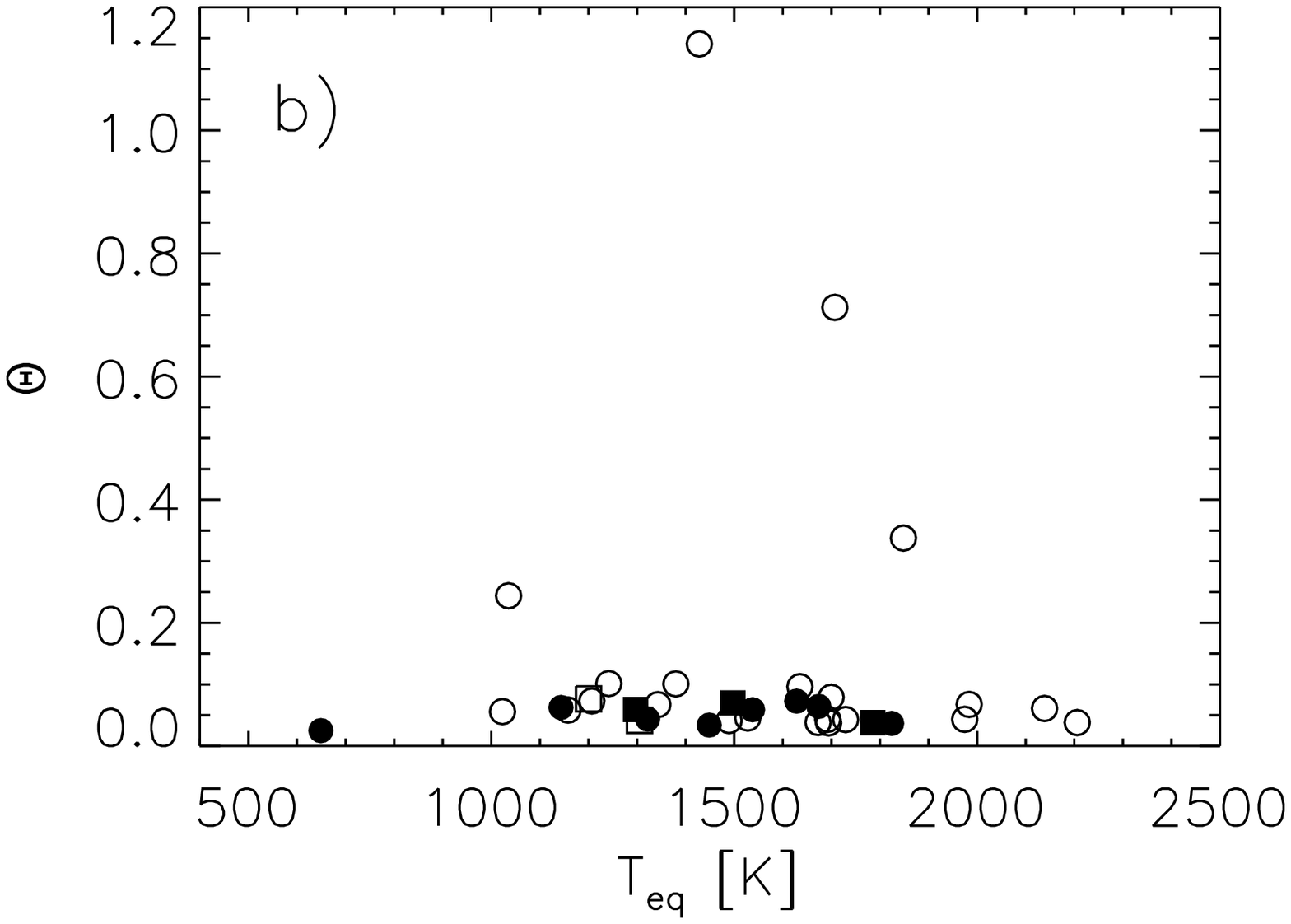}
  \caption{{a)} Safronov number $\Theta$ vs. equilibrium temperature $T_\mathrm{eq}$ for 31 transiting exoplanet systems. Circles show the position of apparently single systems, whereas squares denote known binaries. Objects marked with filled symbols have been observed in the course of the survey presented in this paper. Names in the plot identify the binary transit hosts.
{b)} Same as a) but with a different scale for $\Theta$ showing four additional objects with large Safronov numbers.}
    \label{fig3}
\end{figure*}

From literature studies\footnote{References in http://www.exoplanet.eu/\ }, we could find 35 transiting planets for which Safronov numbers could be calculated. In Figure~\ref{fig3} a) we show the 31 transit planets with Safronov numbers smaller than 0.1.
The distinction between the Class I and Class II planets at $\Theta \approx 0.05$ is still present. Interestringly, three of the 31 exoplanets are located above the former upper limit for Class I at $\Theta\!\le\!0.08$. These three might either form yet another group (``Class\,III''?) or could be seen as an extension of Class I. In Figure~\ref{fig3} b) we show the full range of Safronov numbers for the 35 transit planets, which range up to $\Theta\!\approx\!1.2$. The planets with the highest Safronov numbers are in general the ones with the highest planetary mass relative to the stellar mass.

Of the stellar binaries in our sample, the exoplanets around WASP-2 and TrES-2 belong to Class\,I, whereas the exoplanet TrES-4b belongs to Class\,II. As the update from our calculations is small, the membership of the three objects to Class\,I and II did not change. From the literature we identify two more binaries, HD\,189733 (Class\,I) and HAT-P-1 (Class\,II), which both have larger angular separations than the detection cut-off assumed for the AstraLux sample.

Interestingly, for these five objects, there might also be a correlation between the binary separation and the planet class. The three binaries with projected separations between 110 and 230 Astronomical Units (AU) host Class\,I exoplanets, whereas the wider binaries with their projected separations of $\ge$\,750\,AU\ host Class\,II exoplanets. The presence of a binary companion could very well influence the planet formation process as well as the predominant migration mechanism.

Clearly our sample of five potential stellar companions to transit planet host stars is too small to draw any firm conclusions. Yet it should be worthwhile to explore if this correlation still holds once the binary sample has been expanded.

\section{Summary}
High-resolution imaging of 14 exoplanet hosts stars revealed stellar companions to three of the targets (WASP-2, TrES-2, and TrES-4), the latter two of which have not been known to be multiple before.

{Observations are part of an ongoing imaging survey aiming at characterizing all transit hosts stars with the high-resolution imager {\em AstraLux}. It employs the Lucky Imaging technique which allows spatial resolutions comparable to adaptive optics observations with much smaller overhead and therefore high observing efficiency.}

We present an analysis of the photometry and astrometry obtained from our observations. Together with transit light curve and spectroscopic parameters, planetary and stellar parameters were derived and compared to earlier derivations.

We find the companions to be classified as K7 to M0.5, with a projected separation of 100 to 750\,AU to their F8 to K1 candidate parent stars assuming physically bound companions. While physical binarity has yet to be confirmed through second epoch observations, it is not crucial to most of the conclusions drawn.

Analizing stellar and planetary parameters taking the newly discovered binarity into account we find values to differ from previously derived parameters assuming a single host star: all parameters (e.g. $R_\mathrm{p}/R_\ast$, $a/R_\ast$, $b$, $\rho_\ast$, \dots) have undergone revision and updated values have been shown to differ by up to more than 2$\sigma$, where $\sigma$ is the corresponding uncertainty of each parameter.

A correlation between the Safronov number $\Theta$ and binary separation has been suggested. Wide binaries seem to fall into Class\,II ($\Theta\!\approx\!0.04$) whereas Class\,I ($\Theta\!\approx\!0.07$) seems to host smaller separation binaries in a $\Theta$--$T_\mathrm{eq}$ plot. The very small number of known multiple transit host binaries, however, needs to be increased through future observing in order to confirm or reject this proposal.

\begin{acknowledgements}
The authors would like to thank the referee, Dr. Hans Deeg, for many helpful suggestions and comments which lead to a highly improved paper.
This publication makes use of data products from the Two Micron All Sky Survey, which is a joint project of the University of Massachusetts and the Infrared Processing and Analysis Center/California Institute of Technology, funded by the National Aeronautics and Space Administration and the National Science Foundation.
This research has made use of the SIMBAD database, operated at CDS, Strasbourg, France.
\end{acknowledgements}

\end{document}